\documentclass{aa}

\usepackage{psfig}
\usepackage{graphicx}
\usepackage{natbib}
\usepackage[english]{babel}  
\usepackage{txfonts}
\def\gsim{\;\lower4pt\hbox{${\buildrel\displaystyle >\over\sim}$}\,}
\def\lsim{\;\lower4pt\hbox{${\buildrel\displaystyle <\over\sim}$}\,}

\begin{document}

\title{Shock-cloud interaction in the Vela SNR observed with \emph{XMM-Newton}}

\author{M. Miceli\inst{1} \and F. Bocchino\inst{2} \and A. Maggio\inst{2} \and F. Reale\inst{1}}

\offprints{M. Miceli,\\ \email{miceli@astropa.unipa.it}}

\institute{Dipartimento di Scienze Fisiche ed Astronomiche, Sezione di Astronomia, Universit\`a di Palermo, Piazza del Parlamento 1, 90134 Palermo, Italy
\and 
INAF - Osservatorio Astronomico di Palermo, Piazza del Parlamento 1, 90134 Palermo, Italy
}

\date{Received 27 August 2004 / accepted 8 June 2005}

\authorrunning{M. Miceli et al.}
\titlerunning{X-ray observation of the Vela FilD}

\abstract{We analyzed an \emph{XMM-Newton} EPIC observation of a bright knot, named FilD, in the northern rim of the Vela SNR, where the shock has encountered a cloud. The good combination of sensitivity, spectral, and spatial resolution allowed us to describe the internal structure of the observed ISM clouds and to obtain estimates of their temperature, density, O, Ne, and Fe abundances, and of their extension along the line of sight. We also examined the interaction of the shock with the FilD knot and estimated that the time elapsed from the shock impact is about one cloud crushing time. Our analysis allowed us to conclude that the observed X-ray emission is best explained by the propagation of transmitted shocks through ISM inhomogeneities with an inward increasing density profile. We found that the interaction with the shock determines the heating of the inner part of the clouds and the evaporation of the outer part.

\keywords{X-rays: ISM --  ISM: supernova remnants -- ISM: clouds -- ISM: kinematics and dynamics -- ISM: individual object: Vela SNR}}

\maketitle

\section{Introduction}
\label{Introduction}

Supernova remnants (SNRs) play a leading role in settling the mass distribution and the energy exchanges in the interstellar medium (ISM) and, according to \citet{cs74}, they account for the heating of about 1/10 of the volume of the interstellar medium of our galaxy. During their expansion, SNRs heat, compress, and ionize the ISM and determine the destruction of the interstellar dust grains and the consequent release of heavy elements in the interstellar gas.

The inhomogeneities of the ambient medium influence the direction and the speed of the SNRs expansion. The optical and X-ray emissions of middle-aged SNRs arise from the interaction between the shock and the ISM clouds: optical filaments arise from slow radiative shocks propagating in a dense ($n\sim10$ cm$^{-3}$) and cool ($T\sim10^{4}-10^{5}$ K) medium, while X-ray emission originates from hotter and less dense regions. However, the details of the heating processes are still not well understood.

Three different physical scenarios have been proposed to explain the clumpy emission of the SNRs.
A first scenario associates the optical emission to a transmitted shock travelling through dense clouds and explains the X-ray emission with the evaporation of the inhomogeneities engulfed by the main blast wave. This model has been applied in the past to observations of SNRs (\citealt{fbk82}, \citealt{ckm85}, \citealt{chp97}) and, more recently, to describe the composite SNR G327.1-1.1 (\citealt{bb03}).

A second class of models points out the importance of the overpressured regions in the post-shock flows and associates the X-ray emission to the inter-cloud medium heated by reflected shocks. Hydrodynamical simulations show that the cloud density contrast drives reverse reflected shocks in the inter-cloud medium (see e.g. \citealt{sn92}). According to this model the X-ray emission is localized ``behind'' the optical emission, which is associated with the shock transmitted into the inhomogeneities. This scenario has been extensively adopted to explain the optical and X-ray emission of the Cygnus Loop (\citealt{glh95}, \citealt{lgh96}, \citealt{mt01}).

A third interpretation is based on the propagation of transmitted shocks through inhomogeneities with an inwards increasing density profile. These shocks can account for the observed X-ray emission  when the cloud contrast is low and/or when the primary blast wave shock is fast. This scenario was proposed by \citet{bms00} to explain the optical and X-ray emission of two regions of the Vela SNR in which the main blast wave
seems to interact with a small isolated cloud (FilD region) and with a cavity wall (FilE region), and has been recently applied by \citet{pfr02} and \citet{lgw03} for two regions of the Cygnus Loop.

The instrumental performances of past generation X-ray satellites did not always allow discrimination between these models or an unambiguous description of the emission mechanism in middle-aged SNRs.

The Vela SNR is the nearest object of its kind, so it represents a privileged laboratory for the study of the shock-cloud interaction.
The distance of the Vela SNR is $\sim 280$ pc (\citealt{bms99}, \citealt{csd99}) and at this distance one arcmin corresponds to only 0.08 pc. The age of this SNR has been estimated to be $\sim 10000$ yr, in good agreement with the characteristic age ($\sim 11200$ yr, \citealt{tml93}) of the pulsar PSR B0833-45, which is associated with the remnant (\citealt{wp80} and \citealt{ws88}). 

The northern rim of the Vela SNR presents all the characteristic features of a complex interaction between the blast wave shock and several inhomogeneities of the ambient medium, as pointed out by the ROSAT observations of FilD and of FilE (\citealt{bms94,bms97,bms99,bms00}). Analysis of the X-ray emission of these regions showed that the observed spectra are described significantly better by a 2-component optically thin thermal plasma in equilibrium of ionization than by a single-temperature model or a non-equilibrium of ionization (NEI) model. As for FilD, the two components with temperatures $1.10\pm 0.06\times 10^{6}$ K and $6^{+6}_{-3}\times 10^{6}$ have been associated with the shocked plasma of the cloud and with the inter-cloud medium, respectively. However the instrumental characteristics of the PSPC detector did not allow us to describe the shock-cloud interaction univocally or to get detailed information about the morphology of the emitting plasma and about its chemical abundances.

Here we present analysis of an \emph{XMM-Newton} EPIC observation of the FilD region. The aim of this work is to exploit the EPIC characteristics to derive a better description of the shock-cloud interaction process in this region of the Vela SNR and to discriminate among the different models described above.
 
The paper is organized as follows: in Sect. \ref{The Data} we present the EPIC data analyzed here, while in Sect. \ref{The data analysis} we describe the morphology of the X-ray emission, identifying the physically different emitting regions, and then present the results of the spatially resolved spectral analysis. In Sect. \ref{Discussion} we study the thermal structure of the source, show the 3D map of the plasma, and discuss the shock-cloud interaction in this region of the Vela SNR. The presence of NEI effects is examined in Sect. \ref{Discussion} and, finally, our conclusions are drawn in Sect. \ref{Summary and conclusions}.

\section{The Data}
\label{The Data}

The \emph{XMM-Newton} observation of Vela FilD we present is a Guaranteed Time Observation, performed with the EPIC MOS (\citealt{taa01}) cameras and the EPIC pn (\citealt{sbd01}) camera on 9 November 2001 and with pointing coordinates $\alpha$ $(2000)=8^{h}35^{m}44^{s}$, $\delta$ $(2000)=-42^\circ35'29''$. EPIC cameras are CCD detectors with spectral resolution $E/\Delta E\sim 20-50$ in the $0.1-10$ keV band. The field of view (FOV) has a diameter of $\sim 30'$, and the pixel size is $1.1''$ (MOS) and $4.1''$ (pn), while the mirror PSF width is $6''-15''$ FWHM--HEW. The \emph{XMM-Newton} mirror assembly is subject to vignetting, so there is a decrease in effective area with increasing in off-axis angle.

Table \ref{tab:data} summarizes the main information about the data.
\begin{center}
\begin{table}[htb!]
\begin{center}
\caption{Relevant information about the data.}
\begin{tabular}{lccccc} 
\hline\hline
CAMERA &  $t_{exp}$ (ksec)$^{I}$  & Mode              & Filter & $N_{ph}^{II}$  \\ \hline
pn      & 26.8/11.5              & Ext. Full Frame    & medium   &  291115 \\
MOS1    &   31.1/25.2            & Large Window       &   medium & 203113  \\
MOS2    &   31.1/25.2           & Large Window        &   medium & 207807 \\
\hline\hline
\multicolumn{5}{l}{\footnotesize{I. Unscreened/Screened exposure time.}} \\
\multicolumn{5}{l}{\footnotesize{II. Number of events in the whole field of view ($0.3-10$ keV).}} \\
\label{tab:data}
\end{tabular}
\end{center}
\end{table}
\end{center}
 
\subsection{Data processing}
\label{Data Processing}

Data have been processed with the Standard Analysis System (SAS V5.4.1) pipeline. We further screened the data by excluding some residual hot pixels (especially from the MOS1 observation) and by removing the time intervals which showed anomalous background enhancements (probably due to soft protons). In particular, we extracted the background light curve with a bin size of 10 s, at energy $>10$ keV and we eliminated time intervals with high counts ($>30$ for pn data and $>15$ for MOS1 and MOS2 data). Table \ref{tab:data} shows the unscreened and screened exposure time $t_{exp}$; the pn observation is strongly contaminated by proton flares ($\sim\,57\,\%$ of the total exposure).

Spectral analysis was performed using XSPEC. The ancillary response files (arf) were generated with the SAS $ARFGEN$ task, and the event files were processed using the $EVIGWEIGHT$ task (\citealt{ana01}) to correct vignetting effects. We summed the MOS1 and MOS2 spectra averaging the arf files (and summing the exposure times) and using the MOS1 response file for the analysis of the summed spectrum. The fittings were performed simultaneously on the pn and MOS-summed spectra. All the reported errors, calculated according to \citet{lmb76}, are at 90\% confidence level.

\subsection{Background subtraction}
\label{Background Subtraction}

We removed the X-ray sources present in the EPIC field of view and not correlated with the Vela SNR. Detection of these sources was performed on the summed data set (MOS1$+$MOS2$+$pn), using the Wavelet Transform algorithm $PWXDETECT$ (\citealt{dmm97a,dmm97b}). In this way we found position, extension, and X-ray flux of 21 sources; 10 among them have an optical counterpart in the USNO-A2.0 archive. All the detected sources have a significant signal only in the energy range $1-10$ keV, so we subtracted their contribution in this band.
Since the Vela emission covers the entire FOV, it was not possible to isolate a region where the local background could be extracted.

The contribution of the incident photon X-ray background was derived from the EPIC public observation $0106260201$, whose FOV is wholly outside the Vela SNR shell, and pointed toward the source RXJ0806.4-4123, which is only $\sim 5^\circ$ distant from our observation. In the $0.3-2$ keV band (where our analysis is focussed) the count rate contribution of this background is between $\sim 5$ $\%$ and $\sim 15$ $\%$ of the total signal. In particular, in a $4''$ bin, the mean $0.3-2$ keV MOS count rate of the photon background is $R\sim 8.6\times 10^{-6}$ s$^{-1}$ with a mean photon energy $\overline{E}\sim 670$ eV; we used these values to compute the background subtracted $\overline{E}$ vs. $R$ plot described in Sect. \ref{E-rate relation}. We also used this observation to extract a pn and a MOS background spectrum for the spectral analysis. However, as shown in Appendix A, the best fit parameters are not significantly influenced by the choice of the background. 

As for the non-photon background, which is probably associated with a residual flux of soft protons, we found that above 1 keV the non-photon background amounts to $>95\%$ of the total signal for the pn camera and $>70\%$, for the two MOS cameras and that: i) In the $0.3-0.5$ keV band the non-photon background contribution is too low to be estimated; ii) in the $0.5-1$ keV band it amounts to $2-12\,\%$ of the total signal for the pn camera, and iii) in the $0.3-2$ keV band it is negligible.

\section{The data analysis}
\label{The data analysis}

\subsection{The region morphology}
\label{The region morphology}

To reduce statistical fluctuations in the images, we implemented a weighted average of the pn, MOS1, and MOS2 count rate maps in three different bands: $0.3-0.5$ keV, $0.5-1$ keV, and $0.3-2$ keV. The first two bands were chosen to understand how the emission is distributed in the FOV at different energies, while the image in the $0.3-2$ keV band allowed us to make a comparison between the EPIC and the PSPC observations. 
All images are MOS-equivalent, and the weight of the different cameras were chosen taking the thermal spectrum of the source into account.
We estimated that in all these images the relative statistical error of the count rate in each bin is $<18\%$.
   
Fig. \ref{fig:rate03-05-1} shows the count rate images (vignetting corrected) in the $0.3-0.5$ keV band and $0.5-1$ keV band; these images are adaptively smoothed with $\sigma_{min}=4''$ and $\sigma_{max}=24''$. North is up and East is on the left. Notice that, assuming a distance of 280 pc (\citealt{bms99}), the diameter of the EPIC FOV corresponds to an extension of $\sim 2.4$ pc.
There are striking differences in the morphology of the emission in these two bands. The FilD complex, in the middle of the FOV, is clearly visible in the soft band (\emph{left panel} Fig. \ref{fig:rate03-05-1}), and its internal structure is resolved in good detail with \emph{XMM-Newton}. FilD looks composed by two bright parts (A and B in Fig. \ref{fig:rate03-05-1}).
South of FilD there is a wide region, labelled C, that extends westwards. As shown in the {ROSAT} observation RP500015 (\citealt{bms99}, Fig. 2b), region C stretches beyond the EPIC FOV and is probably part of a very large structure. FilD is instead barely visible in the hard band (\emph{right panel}), where another large structure is present. This bright structure, named RegNE, has a slightly higher count rate than FilD in the $0.3-2$ keV band (Fig. \ref{fig:EPIC03-2_avgE}) and was not visible in the \emph{ROSAT} PSPC observation.

It is interesting to note how EPIC data relate to the optical observations of the FilD region. Figure \ref{fig:ottico_0305} shows a color composite optical image of FilD (from \citealt{bms00}), where we have superimposed the X-ray contour levels of the EPIC observation in the $0.3-0.5$ keV band. The FOV of this observation is indicated by the white square in the left panel of Fig. \ref{fig:EPIC03-2_avgE}.

In the southwestern part of the image there is a clear relationship between optical and X-ray emission: maximum X-ray emission contour levels, in fact, are just outside the optical filament which follows a path that ``grazes'', without crossing, the two FilD regions with the highest X-ray surface brightness (A and B in Fig. \ref{fig:rate03-05-1}). 
This kind of relationship between optical and X-ray emission, also observed by \citet{pfr02} in the southwestern region of the Cygnus Loop, suggests that a transmitted shock is propagating through a cloud with an inwardly increasing density profile, generating optical emission in the inner and denser part of the cloud and X-ray emission in the outer part.

\subsection{Analysis of the photon energy map}
\label{E-rate relation}

The \emph{XMM-Newton} data allowed us to study the distribution of photon energies across the FOV and to produce a high resolution mean photon energy ($\overline{E}$) map, shown in Fig. \ref{fig:EPIC03-2_avgE}, right panel. To realize this map, we considered the MOS1 and MOS2 event files in the $0.3-2$ keV band; a bin size of $10''$ allowed us to collect more than 10 counts per pixel everywhere. For each pixel, we calculated the mean photon energy and then we smoothed the map with a gaussian distribution with $\sigma=20''$  (the error on each value of $\overline{E}$ is $\sim 10$ eV). The map clearly shows that $\overline{E}$ has a minimum in the FilD region and a maximum in RegNE.

The large inhomogeneities in the count rate and mean energy maps suggest that the blast wave shock propagates in a complex environment, generating different thermal conditions in the shocked plasma. To get information about the distribution of the physical parameters of the plasma and about the geometrical morphology of the inhomogeneities, we investigated the relation between the mean photon energy in a given pixel and the corresponding count rate $R$.

Figure \ref{fig:avge-rate} shows the observed $\overline{E}$ vs. $R$ relation (background subtracted), where each point corresponds to a $10''$ pixel of the mean energy map.
The global trend is a decreasing one, as expected in the case of pressure equilibrium (\citealt{bmm04}). It is interesting to note how the different structures present in the FOV are disposed in the graph. The black points correspond to the photons emitted from FilD and an adjacent region; they clearly identify a very narrow strip, suggesting that FilD and the surrounding fainter regions are part of a single structure. We can then assume that FilD is a largely isobaric structure formed by  components with different temperatures and/or extensions along the line of sight. The red points, corresponding to the southeastern region, distinctly identify a different flatter trend, with a higher $\overline{E}$, hence a higher mean temperature along the line of sight and a higher value of pressure and/or volume (\citealt{bmm04}). This region may therefore be physically different from FilD. Finally, the events detected in the northeastern region of the FOV (blue) have the highest mean energy and the decreasing trend is less marked. These points do not overlap with the points either of FilD or of the southern region, suggesting that we are observing three physically different ISM inhomogeneities.

Note that a plasma heated by a reflected shock would yield an increasing trend in the $\overline{E}$ vs. $R$ scatter-plot. In fact, the hotter plasma behind the reflected shock front is also more compressed since it was shocked twice.
Therefore, in the reflected shock case, the plasma with the higher temperature should also have a higher emission measure. Thus the $\overline{E}$ vs. $R$ trend, visible in Fig. \ref{fig:avge-rate}, suggests that we are observing ISM clouds heated by a transmitted shock.

\subsection{Spectral analysis}
\label{Spectral Analysis}

The analysis above allows us to define spatial regions in the FOV with homogeneous physical properties across the plane of the image. We can now perform a spatially resolved spectral analysis on them to obtain the physical parameters. We analyzed the spectra extracted from the 16 subregions indicated in the right panel of Fig. \ref{fig:EPIC03-2_avgE}. These subregions cover all the significant structures in the FOV, identified on the basis of the mean photon energy map of Sect. \ref{E-rate relation}, and their shape and size were chosen so as to have enough photon statistics ($>2000$ counts per subregion) and very small fluctuations of the mean photon energy ($<4\%$), which is almost uniform in each subregion. In this way we can probe the different spectral features between regions whose extension is only a few arcminutes and which do not have internal spectral inhomogeneities \emph{across the plane of the image}. Subregions $1-6$ are located on FilD; subregions $7-12$ cover the south western structure; subregion 13 is in the ``transition zone'' between FilD and RegNE which is covered by subregions 14 and 15; subregion 16 corresponds to the northern area with very low surface brightness.

All the extracted spectra are described well by two MEKAL components of an optically-thin thermal plasma in collisional ionization equilibrium, CIE (\citealt{mgv85}, \citealt{mlv86}, \citealt{log95}), with non-solar Ne and Fe abundances and modified by the interstellar absorption of \citet{mm83}. Figure \ref{fig:region7} shows a representative spectrum, extracted from subregion 7, with its best fit model.

The elements whose lines mainly contribute to the observed spectra are O, Ne, and Fe. In particular in all spectra the K-shell line complexes of O VII (at $0.56$ keV), O VIII (at $0.65$ keV), and Ne IX (at $0.92$ keV) are visible. As for the O abundance, we found in each subregion $O/O_\odot=0.9\div 1.1$ so we tied this abundance to 1. The blending between the Fe XVII L-lines and the Ne IX lines generates an entanglement between the best fit values of the abundances of Ne and Fe, and the statistics were not enough to discriminate between these two values in most of the regions. To overcome this problem we focussed our attention on the subregions 14, 15, and 16 where the signal to noise ratio is best at energies of $\sim 0.9$ keV. In these subregions the mean value of the ratio $(Ne/Ne_\odot)/(Fe/Fe_\odot)$ is 4.4 and we assumed that this ratio remains the same in the other subregions. Therefore, in all the subregions, we left the model Fe abundance free and linked the Ne abundance to it, so as to have $(Ne/Ne_\odot)/(Fe/Fe_\odot)=4.4$. 

It was not possible to obtain a good estimate of the $N_{H}$ parameter (we found $N_{H}\simeq 1.5^{+3.5}_{-1.5}\times 10^{20}$ cm$^{-2}$), so it was fixed to $1\times 10^{20}$ cm$^{-2}$, which is a typical value in this part of the Vela SNR (\citealt{bms94}, \citealt{la00}, and \citealt{kgh85}).

Figures \ref{fig:Ne}$-$\ref{fig:N1-N2} describe in detail the findings of our analysis. The achieved results can be summarized as follows:
\begin{enumerate}
\item In the subregions 14 and 15, spectra can be modelled with one component.
\item The temperatures $T_{I}$ and $T_{II}$ of the two components are both rather uniform in the FOV (Fig. \ref{fig:T1T2}) with mean values (between the subregions): $\overline{T}_{I}=1.14\pm 0.02\times 10^{6}$ K (cooler component) and $\overline{T}_{II}=3.0\pm 0.1\times 10^{6}$ K (hotter component).
\item The emission measures of both components have large variations, as shown in Fig. \ref{fig:N1-N2}.
\item A third MEKAL component in the model did not improve the description of the data.
\end{enumerate}

\subsection{Ne and Fe abundances}
\label{Fe and Ne abundances}

Figure \ref{fig:Ne} shows the best-fit values of $Ne/Ne_{\odot}$ in the 16 subregions versus the mean MOS count rate (for a bin size of $10''$) in the corresponding subregion. The Fe abundances can be deduced from the condition $(Ne/Ne_\odot)/(Fe/Fe_\odot)=4.4$. 

In 10 subregions we found Ne abundances significantly higher than cosmic value, with an average $\overline{Ne}/Ne_\odot=1.7\pm 0.2$ over all the subregions. Typically the Ne overabundance is associated with overabundances of O, Mg, Si (\citealt{ssh01}) and Fe (\citealt{mta01}) and is a clear signature of the ejecta, but this is not the case, since Fe is underabundant and we found solar O abundances. Similar values of the Ne abundance in the Vela SNR were found by \citet{pse02}, who showed that the Vela Bullet D X-ray emission can be described with a non-equilibrium of ionization model (which seems to exclude that Bullet D is a fragment of ejecta) with $Ne/Ne_\odot \approx 2$. Moreover, in a small region of the Cygnus Loop, \citealt{mt01} found analogous O, Ne, and Fe abundances  ($Ne/Ne_\odot=1.7^{+0.9}_{-0.3}$, $O/O_\odot=0.8^{+0.3}_{-0.2}$ and $Fe/Fe_\odot=0.50^{+0.22}_{-0.04}$).

In all the analyzed subregions the Fe is underabundant with an average value $\overline{Fe}/Fe_\odot =0.39\pm 0.05$. According to \citet{vrd94} (VRD), the fraction of Fe locked up in the pre-shock ISM grains is $0.98$, so in the X-ray emitting plasma at the shock front we have $Fe/Fe_\odot=0.02$. VRD used a numerical model to describe the evolution of the grain destruction which generates the release of Fe in the interstellar gas in the post-shock region. However, assuming the typical values of this zone of the Vela shell, shock inter-cloud velocity $v_{ic}=700$ km/s, and inter-cloud density $n_{ic}=0.05$ cm$^{-3}$ (Sect. \ref{Cloud crushing time}), this model predicts that the observed Fe abundance can be obtained only after $\sim20000$ yr, i. e. about two times the age of the Vela SNR. 

In conclusion, it is difficult to explain the observed Ne and Fe abundances, and it is necessary to obtain more information about the value and the spatial distribution of the abundances of these elements in the Vela SNR.
 
\section{The physical interpretation}
\label{Discussion}

\subsection{Thermal structure of the region}
\label{Thermal structure of the source}

As we pointed out in the previous section, the observed spectra are described well by two thermal components at $\sim 10^{6}$ K and $\sim 3\times 10^{6}$ K. Since in each subregion the mean photon energy is almost uniform (maximum variations of $\sim 25-30$ eV), two distinct components necessarily indicates two different thermal phases of the interstellar medium intercepted \emph{along the line of sight}. Since the emission measures per unit area are highly inhomogeneous, we can conclude that the EPIC data reveal for the first time that both these phases are associated with the ISM inhomogeneities (therefore, in what follows, we will use the term ``clouds'' to indicate these hot, X-ray emitting structures). This is at variance with ROSAT data, which suggested (\citealt{bms99}) an association of the hotter thermal component with the inter-cloud medium (the tenuous and almost uniform ISM phase which surrounds the clouds). Clouds can so be described as two-temperature structures.

It is also reasonable to assume that the temperature is uniquely determined by the local density, because the shock equations (\citealt{ll59}) predict that the temperature increase of the shocked plasma is proportional to the square of the Mach number $M$ of the shock, which changes with the local density $n$ ($M\propto n^{-1/2}$). We conclude that, since the temperature of each component is uniform in the FOV, the density of each phase is also similar for all the observed clouds. Therefore, differences in surface brightness imply differences in the extension of the two phases along the line of sight. The two components of the X-ray emission can be related to two different parts of the ISM clouds with different densities: a core, which corresponds to the cooler and denser component, and a hotter and less dense corona, which reasonably surrounds the core.

Figures \ref{fig:T1T2} and \ref{fig:N1-N2} show the best-fit temperatures and emission measures versus the mean MOS count rate (for a bin size of $10''$) in the corresponding subregion.
These results allow us to conclude that the trend in the $\overline{E}$ vs. $R$ scatter plot of Fig. \ref{fig:avge-rate} is essentially determined by the (relative) dimensions of the cores and of the coronae.

\subsection{Density of the cores and of the coronae}
\label{density of cores and coronae} 

An upper limit to the density of the cloud cores (and hence of the coronae) is, obviously, the density $n_{opt}$ of the optical filament of Fig. \ref{fig:ottico_0305}. The multi-wavelength study of the FilD region performed by \citet{bms00} has shown that $n_{opt}=3.2-8.2$ cm$^{-3}$, so we infer that the density of the X-ray emitting plasma must be $\la 5$ cm$^{-3}$.

Considering the lower limits (at 90\% confidence level) of the emission measure (shown in Fig. \ref{fig:N1-N2}) and an upper limit for the extension along the line of sight, $L_{max}$, we can get an estimate of the lower limits of $n_{I}$ and $n_{II}$. The shocked inhomogeneities cannot extend beyond the Vela SNR, so the length of the chord intercepted by the Vela shell along the line of sight represents the maximum extension $L_{max}$ of the clouds in this direction. If we approximate the Vela shell as a sphere with radius $r_{Vela}\approx5.06\times10^{19}$ cm, in FilD $L_{max}\approx 5.8\times10^{19}$ cm and $n_{I}>0.32$ cm$^{-3}$, while, in RegNE, $L_{max}\approx 5.1\times10^{19}$ cm and $n_{II}>0.14$ cm$^{-3}$.

On the other hand, we may assume that FilD has an average extension along the line of sight comparable to its size in the plane of the image ($L_{FilD}\sim2\times 10^{18}$ cm) and, in this way, we obtain a best guess value for $n_{I}\sim 1.5$ cm$^{-3}$. In the same way, setting $L_{RegNE}\sim3\times 10^{18}$ cm, we obtain a best guess value for $n_{II}\sim 0.55$ cm$^{-3}$. Notice that these values indicate pressure equilibrium between cores and coronae.

An independent estimate of a lower limit of $n_{I}$ can be obtained from the evidence that the cloud cores have reached the collisional ionization equilibrium. Modelling the cooler component of the spectra with the PSHOCK non-equilibrium model (\citealt{blr01}), we obtained best-fit values of $\tau_{NEI}\approx 2\times 10^{13}$ s/cm$^{-3}$ with a lower limit (for subregion 4) of $2\times 10^{11}$ s/cm$^{-3}$, so:
\begin{equation}
\label{eq:tau}
\tau_{NEI}=\int_{t_{FilD}}^{t_{Vela}} n_{I}(t)dt=\overline{n}_{I}(t_{Vela}-t_{FilD})>2\times10^{11}\hspace{0.18cm}\hspace {0.22cm}{\rm s/cm}^{3}
\end{equation}
where $(t_{Vela}-t_{FilD})$ is the elapsed time since the shock reached FilD, and $\overline{n}_{I}$ is the average density in this period. Assuming the Sedov model describes the evolution of the shock in the inter-cloud medium, we have:
\begin{equation}
\label{eq:tempo}
t_{Vela}-t_{FilD}=\int_{r_{FilD}}^{r_{Vela}}\frac{dr}{v(r)}=t_{Vela}\left[1-\left(\frac{r_{FilD}}{r_{Vela}}\right)^{5/2}\right]\,\,.
\end{equation}
If FilD and the center of the Vela shell lie on a same plane, normal to the line of sight, $r_{FilD}\approx 4.1\times10^{19}$ cm , therefore $t_{Vela}-t_{FilD}\approx 4500$ yr and, from Eq. (\ref{eq:tau}), $\overline{n}_{I}>1.4$ cm$^{-3}$.
Conversely, by putting $\overline{n}_{I}=5$ cm$^{-3}$ we can also obtain a lower limit for $t_{Vela}-t_{FilD}$, in this way we infer that FilD was shocked at least 1300 yr ago.

A similar useful estimate of a lower limit for $n_{II}$ was not possible, because in the coronae we do not have a univocal signature of CIE and the hotter component can also be described with a NEI model, as we show in Sect. \ref{NEI effects}.
Our results can be summarized as follows:

\begin{displaymath}
\textbf{Cores:}\hspace{0.3cm}\left\{ \begin{array}{ll}
                 n_{I}\approx1.5\,\,(0.32\div5) \hspace {0.5cm}{\rm (cm^{-3})} \\
                 \overline{n}_{I} \ga 1.4 \hspace{0.5cm} {\rm (cm^{-3})}
                  \end{array}
          \right.
\end{displaymath}

\begin{displaymath}
\textbf{Coronae:}\hspace{0.3cm} \hspace{0.06cm} n_{II}\approx 0.55\,\,(0.14\div 5) \hspace{0.5cm} {\rm (cm^{-3})}
\end{displaymath}

\subsection{3D map of the plasma}
\label{3-D map of the plasma}

Merging the results of the spectral analysis with the related images, we derived the three-dimensional structure of the observed ISM clouds with resolution down to less than $0.1$ pc. These maps are based on the assumption that the observed X-ray emission at any position in the field of view can be described with two-component thermal models having the same temperatures values, but different emission measure weights. The two temperatures were computed as mean values over all the spectral subregions presented in Sect. \ref{Spectral Analysis}, also assuming the same Ne/Fe ratio anywhere. This approach allows us to derive maps of the extensions along the line of sight of the cores ($L_{I}$) and of the coronae ($L_{II}$)

In each pixel, the MOS count rates $R_{s}$, in the $0.3-0.5$ keV band, and $R_{h}$, between $0.5$ keV and $1$ keV are:
\begin{equation}
\label{eq:paperoneSFT}
R_{s}=P^{MOS}_{s}(\overline{T}_{I})\frac{n^{2}_{I}L_{I}\Omega}{4\pi}+P^{MOS}_{s}(\overline{T}_{II})\frac{n^{2}_{II}L_{II}\Omega}{4\pi}
\end{equation}
\begin{equation}
\label{eq:paperoneMD}
R_{h}=P^{MOS}_{h}(\overline{T}_{I})\frac{n^{2}_{I}L_{I}\Omega}{4\pi}+P^{MOS}_{h}(\overline{T}_{II})\frac{n^{2}_{II}V_{II}\Omega}{4\pi}
\end{equation}
where $P^{MOS}_{s,h}(T)$ is the function of radiative losses per emission measure folded with the \emph{XMM-Newton} MOS response (in the corresponding energy band) and $\Omega$ is the solid angle subtended by a pixel. The $P^{MOS}_{s,h}(T)$ values were calculated with the simulation of a MOS observation (in the soft and in the hard band) of an isothermal plasma with temperature $\overline{T}_{I,II}$, abundances $\overline{Fe}/Fe_\odot$ and $\overline{Ne}/Ne_\odot$ and with $(n^{2}_{I}L_{I}\Omega/4\pi)=1$. In this way we found: $P^{MOS}_{s}(\overline{T}_{I})=20.08$ cm$^{5}$ s$^{-1}$; $P^{MOS}_{h}(\overline{T}_{I})=14.21$ cm$^{5}$ s$^{-1}$; $P^{MOS}_{s}(\overline{T}_{II})=53.53$ cm$^{5}$ s$^{-1}$; and $P^{MOS}_{h}(\overline{T}_{II})=191.02$ cm$^{5}$ s$^{-1}$.

Solving the system of Eqs. (\ref{eq:paperoneSFT}) and (\ref{eq:paperoneMD}) for $n^{2}_{I,II}L_{I,II}$ we obtain:
\begin{equation}
\label{eq:n2LSFT}
n^{2}_{I}L_{I}=\left(\frac{4\pi}{\Omega}\right)\frac{R_{h}P^{MOS}_{s}(\overline{T}_{II})-R_{s}P^{MOS}_{h}(\overline{T}_{II})}{P^{MOS}_{s}(\overline{T}_{II})P^{MOS}_{h}(\overline{T}_{I})-P^{MOS}_{s}(\overline{T}_{I})P^{MOS}_{h}(\overline{T}_{II})}
\end{equation}
\begin{equation}
\label{eq:n2LMD}
n^{2}_{II}L_{II}=\left(\frac{4\pi}{\Omega}\right)\frac{R_{s}P^{MOS}_{h}(\overline{T}_{I})-R_{h}P^{MOS}_{s}(\overline{T}_{I})}{P^{MOS}_{s}(T_{II})P^{MOS}_{h}(\overline{T}_{I})-P^{MOS}_{s}(\overline{T}_{I})P^{MOS}_{h}(\overline{T}_{II})}\,\,\,.
\end{equation}
Notice that Eqs. (\ref{eq:paperoneSFT}) and (\ref{eq:paperoneMD}) hold only if the count-rate in each pixel does not depend on the count rate in the adjacent pixels, so the angular extension of the pixels has to be larger than that of the PSF. For this reason we realized two MOS count-rate images (in the $0.3-0.5$ keV and $0.5-1$ keV bands respectively) by rebinning with a pixel-size of $16''$. We then applied Eqs. (\ref{eq:n2LSFT}) and (\ref{eq:n2LMD}) at each pixel of these images and thereby obtained a high resolution map of $n^{2}L$ for the two components. From this map it is possible to know the emission measure of the plasma with a spatial resolution $\la0.05$ pc. It is important to remark that the results obtained from Eqs. (\ref{eq:n2LSFT}) and (\ref{eq:n2LMD}) are consistent with the findings of the spectral analysis; for example, in subregion 14 (where $T_{II}$ is significantly different from $\overline{T}_{II}$) the best fit value of $n^{2}_{II}L_{II}$ is $1.13\pm 0.05\times 10^{18}$ cm$^{-5}$, while from the $n^{2}L$ map we obtain that, in this subregion, the mean value of $n^{2}_{II}L_{II}$ is $\sim1.09\times 10^{18}$ cm$^{-5}$.

Assuming our best values for the density of the cloud cores and of the cloud coronae, $n_{I}=1.5$ cm$^{-3}$ and $n_{II}=0.55$ cm$^{-3}$ (Sect. \ref{density of cores and coronae}), respectively, we derived the extension along the line of sight of the two components for each pixel of the $n^{2}L$ map; Figs. \ref{fig:3D1} and \ref{fig:3D2} show the 3D structure of the cores (in white) and of the coronae (in transparent blue), presented from two different angle-shots.

In the map the large core of FilD is surrounded by a very thin corona, the two peaks correspond to the two bright parts of the FilD (A and B in Fig. \ref{fig:rate03-05-1} and subregions 2 and 4 of Fig. \ref{fig:EPIC03-2_avgE}), and the optical filament of Fig \ref{fig:ottico_0305} lies between them. In RegNE the core is nearly absent and the corona is very thick. It is possible that the core of RegNE lies outside of the EPIC FOV, but we can also suppose that this cloud is just behind the front shock and that its core has not yet been reached by the shock (see also Sect. \ref{NEI effects}). Compared to FilD, the southwestern structure has a smaller core and a thicker corona.

We can also derive an estimate of the mass of FilD. The dimensions of its core are $\sim 2.4\times 10^{18}$ cm and $\sim 1.1\times 10^{18}$ cm (on the plane of the observation) and $6.8\times 10^{18}/n_{I}^{2}$ cm (along the line of sight), while its corona has about the same extension on the plane of the observation; but it is an order of magnitude less thick, so the mass of FilD, for $n_{I}=1.5$ cm$^{-3}$, is $\sim 10^{31}$ g (the mass of the corona is $\sim 7\times 10^{29}$ g).
The dimensions of RegNE are $\sim 1.4\times 10^{18}$ cm and $\sim 3.3\times 10^{18}$ cm, on the plane of the observation, and $\sim1\times 10^{18}/n_{II}^{2}$ cm, along the line of sight.

Note that with other guesses of the $n_{I}$ value, assuming pressure equilibrium, the extensions along the line of sight of cores and coronae scale with the same factor ($L\propto 1/n^{2}$); hence, the relative size of the inhomogeneities does not change.

It is possible to show that the 3D map is consistent with the hypothesis of pressure equilibrium between cores and coronae. In fact the time-scale for pressure balance is the sonic time-scale. Therefore we expect pressure equilibrium into structures with dimensions comparable to the length $\Lambda$ covered at the sonic speed $c_{s}=(\gamma k_{b}T/\mu)^{1/2}$ in 4500 yr, which is the time elapsed from the shock-impact. $\Lambda_{I}\approx 1.8\times 10^{18}$ cm for the cooler component, and $\Lambda_{II}\approx 2.8\times 10^{18}$ cm for the hotter component; and these values are indeed comparable to the total (core$+$corona) extensions of the observed inhomogeneities, thus supporting the hypothesis of pressure equilibrium.

\subsection{Shock-cloud interaction}
\label{Shock-cloud interaction}

In this section we discuss the dynamics and evolution of the shock-cloud interaction in the observed region in light of our new results.

The results of our analysis allow us to conclude that, since we neither see enhancement of the X-ray emission ``behind'' the optical filament nor inhomogeneities in the plasma temperature and density, the reflected shock model (see Sect. \ref{Introduction}) is not compatible with our data. This result is further supported by the trend of the points in the $\overline{E}$ vs. $R$ scatter-plot, as explained in Sect. \ref{E-rate relation}. 

\subsubsection{Cloud crushing time}
\label{Cloud crushing time}

An important time-scale for evolution of the shock-cloud interaction is the cloud crushing time $t_{cc}$ (\citealt{kcm90}), defined for a spherical cloud with radius $\rho$ as
\begin{equation}
\label{eq:tcc}
t_{cc}=\frac{\rho\chi^{1/2}}{v_{ic}}
\end{equation}
where, for FilD, $\rho\approx1.2\times 10^{18}$ cm and $\chi$ is the ratio of the density of the cloud to that of the inter-cloud medium. Since $v_{ic}/\chi^{1/2}$ is the velocity of the shock into the cloud, this is the time-scale for the transmitted shock to cross the cloud.

To obtain an estimate of $\chi$, we found an upper limit for the inter-cloud density $n_{ic}$. This value was obtained introducing a third hotter MEKAL component in the analysis of the spectra of the subregions 12 and 16, where there are no distinct signatures of the presence of relevant inhomogeneities, and considering the upper limit of its emission measure. In this way we found $n_{ic}\la 0.06$ cm$^{-3}$. Since the density of FilD is $\sim 1.5$ $^{-3}$ (Sect. \ref{density of cores and coronae}), $\chi\approx30$.

We can estimate $v_{ic}$ using the Sedov model, in this case:
\begin{equation}
\label{eq:v_di_r}
v(ic)=\frac{2r_{Vela}^{5/2}}{5t_{Vela}}r_{FilD}^{-3/2}
\end{equation}
where $r_{FilD}>4.1\times 10^{19}$ cm (Sect. \ref{density of cores and coronae}) and $r_{FilD}<5.0\times 10^{19}$ cm. This upper limit is obtained from Eq. (\ref{eq:tau}) with $t_{Vela}-t_{FilD}\ga 1300$ yr and then from Eq. (\ref{eq:tempo}). Therefore $v_{ic}=5.7-7.8\times 10^{7}$ cm/s. In this way we get
\begin{displaymath}
2700<t_{cc}<3700\hspace {0.4cm}{\rm yr}.
\end{displaymath}

We have already pointed out (Sect. \ref{density of cores and coronae}) that FilD was shocked less than 4500 yr ago. Moreover, assuming $n_{I}=1.5$ cm$^{-3}$, we obtain that the time elapsed from the shock impact is about 4200 yr; hence, we conclude that we are observing the shock-cloud interaction at $\ga 1t_{cc}$.
 
\subsubsection{Thermal conduction and radiative cooling}
\label{Thermal conduction and radiative cooling}

The evolution of the shocked ISM clouds, apart from dragging, is governed by competition of radiative cooling and thermal conduction with the surrounding medium, so it is important to estimate the efficiency of both these processes to describe the scenario of the shock-cloud interaction.

The radiative cooling time-scale of an optically thin plasma is defined as:
\begin{equation}
\label{eq:tauRAD}
\tau_{r}=\frac{p}{(\gamma -1)n_{e}n_{H}P(T)}
\end{equation}
where $\gamma$ is the the specific heat ratio.
The values of $P(T)$ in the $0.01-50$ keV band (calculated according to \citealt{mgv85}, \citealt{mlv86}, \citealt{log95}) for the temperature of the cores and of the coronae and for the observed abundances, are $P(\overline{T}_{I})\approx1.0\times 10^{-22}$ and $P(\overline{T}_{II})\approx3.4\times 10^{-22}$ erg s$^{-1}$ cm$^{3}$.

Considering the \citet{spi62} thermal conduction, the thermal conduction time-scale is:
\begin{equation}
\label{eq:tauCOND}
\tau_{c}\approx\frac{7nk}{2(\gamma -1)}\frac{1}{k_{s}T^{5/2}(1/l^{2}_{x}+1/l^{2}_{y}+1/l^{2}_{z})}
\end{equation}
where $l_{i}$ indicates the temperature scale-height in the three dimensions and $k_{s}=5.6\times 10^{-7}$ erg s$^{-1}$ cm$^{-1}$ K$^{-7/2}$. Since in the observed inhomogeneities we detected only two temperatures, we can reasonably approximate the temperature scale-heights with the extensions of the clouds. The dimensions of FilD and of RegNE are reported in Sect. \ref{3-D map of the plasma}; the southwestern structure extends beyond the EPIC FOV, so it is not possible to estimate its dimensions.
 
We estimated $\tau_{r}$ and $\tau_{c}$ for the cores and the coronae of FilD and of RegNE for all the permitted values of the density (Sect. \ref{density of cores and coronae}), assuming pressure equilibrium between cores and coronae. 
The coronae are heated through thermal conduction from the hotter and less dense inter-cloud medium; on the other hand, they lose their thermal energy through radiative losses and thermal conduction towards the cooler cores. Notice that, since the inter-cloud medium is hotter and less dense than the coronae, Eq. (\ref{eq:tauCOND}) predicts that the heating of the coronae is more efficient than their cooling through thermal conduction. 

In FilD, for all the permitted values of density, we found $\tau_{r}\approx 4\times 10^{5}$ yr  and $\tau_{c}\approx 2\times 10^{3}$ yr: the radiative cooling is negligible, so the FilD corona gains thermal energy by means of thermal conduction from the inter-cloud medium. This heating produces the progressive evaporation of the corona. It is possible to estimate the rate of evaporation of the corona using the \citet{mc77} relation for a spherical cloud with radius $\rho_{pc}$ (in pc) :
\begin{equation}
\label{eq:evaporaz}
\frac{dm}{dt}=-2.75\times 10^{4}T_{ic}^{5/2}\rho_{pc}(30/\ln{\Lambda}) \hspace {0.3cm}{\rm g/s}
\end{equation}
where $m$ is the mass of the cloud and $\ln{\Lambda}$ is the Coulomb logarithm (for $T>4.2\times 10^{5} K$, $\ln{\Lambda}=29.7+\ln(n^{-1/2}_{ic}(T_{ic}/10^{6}))$. Assuming $T_{ic}=6\times 10^{6}$ K, $n_{ic}=0.05$ cm$^{-3}$ (Sect. \ref{Cloud crushing time}) and $\rho_{pc}=0.5\times(l_{x}+l_{y}+l_{z})/3$, we obtain an evaporation time-scale $\tau_{ev}=m/(dm/dt)$ of $\sim100$ yr. This means that the FilD corona has almost totally evaporated in the few thousand years elapsed from the shock impact, in agreement with Fig. \ref{fig:3D1} and Fig. \ref{fig:3D2}, where this corona is thinner than the others.

The FilD core is heated through thermal conduction from the surrounding hotter corona. On the other hand, it loses thermal energy through radiative losses and, if we assume that the optical filament lies inside FilD, through thermal conduction toward the optical filament. The radiative and conductive cooling time scale are invariably of the same order ($\tau_{r}/\tau_{c}=0.8\div1.5$) and are $\sim 5\times 10^{4}$ yr ($\sim10^{5}$ yr, for $n_{I}=1.4$ cm$^{-3}$, and $\sim 2\times 10^{4}$ yr, for $n_{I}=5$ cm$^{-3}$). The heating time-scale is shorter and corresponds to the thermal conduction cooling time-scale of the FilD corona ($\sim 2000$ yr). It is possible to estimate whether this core is gaining or losing energy: the heating rate is $H=(3n_{II}kT_{II}V_{II})/\tau_{c}$, where $V_{II}$ is the volume of the FilD corona and $\tau_{c}$ is the conductive time-scale of the corona; the cooling rate is $C\approx(3n_{I}kT_{I}V_{I})/\tau_{r}$, where $V_{I}$ is the volume of the core and $\tau_{r}$ is its radiative cooling time-scale. The ratio $C/H$ is always lower than 0.22, so we conclude that there is a net heating of the FilD core.

Since the characteristic time-scale of energy exchanges between the core and corona are comparable with the sonic time-scale, the hypothesis of pressure equilibrium is self-consistent.

As for RegNE, where the core is nearly absent, for all the permitted values of density, $\tau_{r}\approx 500\tau_{c}$. Hence the radiative cooling is also negligible in this cloud. Using Eq. (\ref{eq:evaporaz}) we estimated that the time necessary for the evaporation of this corona to be completed is $\sim 500$ yr.

\subsection{The physical scenario in FilD}
\label{The general scenario}

As we pointed out in the previous sections, the thermal conduction between the cores and the coronae and especially between the coronae and the inter-cloud medium is a very efficient process. On the other hand, the estimated value of $t_{cc}$ (Sect. \ref{Cloud crushing time}) and the morphology of the optical filament within the FilD core (and into the southwestern structure) indicate that a transmitted shock is propagating through a cloud with an increasing density profile (\citealt{bms00}). The emerging scenario is that the observed X-ray emission comes from the cloud plasma heated ``internally'' by the transmitted shock and ``externally'' by the thermal conduction from the hotter inter-cloud medium. In the inner and denser part of the cloud, the shock is radiative and the plasma cools off through radiative optical emission, while in the cores and in the coronae there is a net heating of the plasma which evaporates in the inter-cloud medium in a few $10^{3}$ yr. Fig. \ref{fig:cloud} is a schematic representation of the structure of the FilD cloud, which summarizes all the ISM phases we identified on the basis of optical and X-rays data; we also report the values of temperature and particle density of each phase.

\subsection{NEI effects}
\label{NEI effects}

In all the extracted spectra, the fittings of the cooler component with the PSHOCK non-equilibrium of ionization model (\citealt{blr01}) revealed that the cores have reached the collisional ionization equilibrium, because the best-fit values of the $\tau_{NEI}$ parameter are a few $10^{13}$ s/cm$^{-3}$ (and their lower limits are always larger than a few $10^{11}$ s/cm$^{-3}$), and the best-fit temperatures and emission measures are the same as those obtained with the MEKAL model.

The statistics of the hotter component are too low to infer significant information on NEI effects in the FilD and in the southwestern coronae, while in RegNE (subregions 14 and 15) the spectra are described well by the non-equilibrium of ionization model with a temperature of $\sim7\times10^{6}$ K. The results obtained in this region with the CIE and NEI conditions are reported in Table \ref{tab:nei}. Assuming NEI conditions for the hotter component, the temperature of RegNE is a bit higher, and comparable to the inter-cloud temperature (\citealt{bms99}). This is at odds with the patchy morphology of the emission of this region, which seems to suggest that RegNE cannot be associated with the inter-cloud medium. 

\begin{center}
\begin{table}[htb!]
\caption{Best-fit parameters of the spectral analysis of subregions 14 and 15 with the MEKAL CIE model and with the PSHOCK NEI model.}
\begin{center}
\begin{tabular}{lcc} \hline\hline
                               &                           &                       \\ 
 MEKAL                         &      subregion 14         &    subregion 15        \\ 
                               &                           &                      \\ \hline
Temperature ($10^{6}$ K)       &   $2.58^{+0.09}_{-0.1}$   &  $2.73^{+0.13}_{-0.1}$ \\ \hline
$n^{2}L$ ($10^{17}$ cm$^{-5}$) &  $11.3^{+0.5}_{-0.5}$     &  $11.2^{+0.7}_{-0.6}$ \\ \hline
$Fe/Fe_\odot$                   &   $0.64^{+0.14}_{-0.11}$    &  $0.54^{+0.13}_{-0.14}$ \\ \hline
$\chi^{2}$(d. o. f.)      &       $146.4(125)$             &   $158.7(134)$     \\ \hline
                               &                           &                   \\ 
 PSHOCK                        &      subregion 14         &     subregion 15    \\
                               &                           &                   \\ \hline
Temperature ($10^{6}$ K)         &    $7^{+2}_{-2}$        &  $7^{+2}_{-3}$ \\ \hline
$n^{2}L$ ($10^{17}$ cm$^{-5}$) &  $2.3^{+1}_{-0.5}$        & $2.5^{+2}_{-0.4}$ \\ \hline
$Fe/Fe_\odot$                   &  $0.49^{+0.07}_{-0.07}$     & $0.50^{+0.09}_{-0.10}$  \\ \hline
$\tau_{NEI}$ (s/cm$^{3}$)      & $6^{+4}_{-2}\times 10^{10}$ & $8^{+5}_{-3}\times 10^{10}$\\ \hline
$\chi^{2}$(d. o. f.)      &      $131.2(124)$               &    $145.6(133)$           \\ \hline\hline
\end{tabular}
\end{center}
\label{tab:nei}
\end{table}
\end{center}

A more detailed study of NEI conditions will require more data; however, we stress that our conclusions are not affected by NEI. In fact, since the temperature of the hotter component is higher than in CIE, the thermal conduction to the cores is more efficient, so the time-scale for the heating of the cores is lower and the evaporation rate of the coronae is higher than in CIE.

However, from preliminary analysis of the archive EPIC observation 0137550901, pointing the Vela SNR at coordinates $\alpha =8^{h}48^{m}54^{s}$ and $\delta=-45^\circ 39'37''$, we found that in that region of the Vela shell the plasma can be described with two thermal components at $\sim 1\times 10^{6}$ K and $3\times 10^{6}$ K, respectively, in equilibrium of ionization. This result indicates that the cloud coronae can be in equilibrium of ionization.

\section{Summary and conclusions}
\label{Summary and conclusions}

Analyzing the EPIC observation of the Vela FilD region, we derived a detailed description of the morphology of the emitting plasma. In particular, the data allowed us to link the optical and X-ray emission of the FilD cloud. The optical filament lies between the two FilD regions with the highest X-ray surface brightness and this correlation, which seems to be present also in the southwestern structure, indicates that the emission originates from ISM inhomogeneities with an inward increasing density profile. 

The study of the relationship between the observed count rate and the mean photon energy allowed us to separate the different inhomogeneities present in the FOV.
The spectral analysis, performed on the spectra extracted from 16 subregions, showed that the plasma is described well by an optically-thin plasma in CIE with two thermal components. The two components are associated with two different phases of the clouds: the cloud core (which corresponds to the cooler component), surrounded by a hotter and less dense corona. The temperature of the cloud cores is uniform in the FOV ($\sim 1.14\pm 0.02\times 10^{6}$ K) and also all the cloud coronae have the same temperature ($3.0\pm 0.1\times 10^{6}$ K). The similarity of the temperatures implies almost uniform densities among the cores and among the coronae. The inhomogeneous X-ray emission is instead determined by large variations of the emission measure of both components. This indicates different volume distributions along the line of sight and allowed us to derive a 3D distribution of the observed plasma for the first time. 

The observed X-ray emission cannot be associated with the inter-cloud medium, and the reflected shock model (\citealt{hrb94}) is not compatible with our data. The ISM inhomogeneities are heated by the transmitted shock which, travelling through the clouds, generates X-ray and optical emission. 

We also addressed the role of thermal conduction and radiative losses in the post-shock evolution of the clouds. We found that thermal conduction is much more efficient than radiative losses, and it causes heating of the clouds. This heating determines the evaporation of the outer layers of the clouds; after about one cloud crushing time the corona of FilD is almost wholly evaporated. 

We also derived the mean abundances, between all subregions, of O ($\overline{O}/O_\odot\approx 1.0$), Fe ($\overline{Fe}/Fe_\odot =0.39\pm 0.05$), and Ne ($\overline{Ne}/Ne_\odot=1.7\pm 0.2$). The Fe abundance can hardly be reconciled with the post-shock dust grain destruction model. We do not have a straightforward explanation for Ne overabundance.

These results allow us to deepen our knowledge of the interaction of an SNR shock with a small ($\la 1 $ pc) isolated inhomogeneity. It is then interesting to verify if our conclusions can also be generalized for other regions of the Vela shell. Other authors have pointed out that physical conditions of the Vela SNR environments are very different, e. g. the southwestern part vs. the northeastern part. For instance, \citet{ns04} noted that the presence of large dense clouds can explain their discovery of a band of strong high excitation C I near the western front of the Vela SNR, where the X-ray emission decreases and many optical filaments are present. Moreover, the $^{12}$CO survey of the Vela SNR presented by \citet{myo01} shows that the large scale spatial correlation between the X-ray emission and the molecular clouds seems to be quite complicated, while \citet{la00} have shown that the temperatures and EMs of the two thermal components necessary to describe the X-ray spectra present significant inhomogeneities across the remnant. We then expect that the physical parameters we found in FilD may not be representative of the entire Vela SNR. In this framework, in a forthcoming paper, we aim to perform a systematic study of small regions, selected on the basis of large surveys, to shed light on the correlation between X-ray emitting plasma and other ISM components (dust, CO , etc.).

Future studies should also address other open problems, like the lack of NEI effects and Fe and Ne abundances. Moreover, to better understand the physics of the shock-cloud interaction, our further investigations will be focussed on the comparison between the observation results and those that will be obtained by performing detailed hydrodynamical simulations.
 
\begin{acknowledgements}
This work was partially supported from the Ministero dell'Istruzione,
dell'Universit\`{a} e della Ricerca and from the Istituto Nazionale di Astrofisica

\end{acknowledgements}

\bibliographystyle{aa}
\bibliography{references}

\newpage

\begin{figure}[htb!]
\caption{Adaptively smoothed count rate images (MOS equivalent) in the $0.3-0.5$ keV band (\emph{left}) and $0.5-1$ keV band (\emph{right}). These images are a weighted average of the pn, MOS1, and MOS2 images. Eight contour levels, equispaced between $0\%$ and $100\%$ (included) of the peak value ($\sim 7.1\times 10^{-5}$ cnt s$^{-1}$ bin$^{-1}$ and $\sim 10^{-4}$ cnt s$^{-1}$ bin$^{-1}$ respectively), are shown; the bin is $4''$.}
\label{fig:rate03-05-1}
\end{figure}

\begin{figure}[htb!]
 \centering
\caption{\emph{Left}: Adaptively smoothed count rate images (MOS equivalent) in the $0.3-2$ keV band (bin size$=4''$); Eight contour levels, equispaced between $0\%$ and $100\%$ (included) of the peak value ($\sim 1.7\times 10^{-4}$ cnt s$^{-1}$ bin$^{-1}$) are shown. The white square indicates the field of view of Fig. \ref{fig:ottico_0305}. \emph{Right}: Mean energy map (smoothed with $\sigma =20''$, bin size$=10''$). In each pixel the local mean photon energy of the photons detected by the MOS cameras between $0.3$ keV and $2$ keV is reported. We superimposed 6 contour levels equispaced between $0.5$ keV and $0.85$ keV. The 16 subregions selected for spectral analysis are also shown.}
\label{fig:EPIC03-2_avgE}
\end{figure}

\begin{figure}[h!]
\caption{Color composite optical image of FilD (from \citealt{bms00}): in the electronic version, the $\,H \alpha$ emission is reported in green, and the [OIII] emission in violet. The $5.7'\times 5.7'$ field of view is indicated by the white square in the left panel of Fig. \ref{fig:EPIC03-2_avgE}. We superimposed (in red) X-ray contour levels derived from EPIC observation in the $0.3-05$ keV band.}
\label{fig:ottico_0305}
\end{figure}

\begin{figure}[htb!]
\caption{$\overline{E}$ vs. count rate scatter plot for the pixel of the left panel of Fig. \ref{fig:EPIC03-2_avgE}. In the electronic version the colors of the points codify the three different regions indicated in the field of view. Errors in $\overline{E}\sim 10$ eV, bin size$=10''$.}
\label{fig:avge-rate}
\end{figure}

\begin{figure}[htb!]
\caption{pn (upper) and MOS-summed (lower) spectra extracted from subregion 7 with the corresponding best fit model and residuals.}
\label{fig:region7}
\end{figure}

\begin{figure}[htb!]
\caption{Best fit $Ne$ abundances in units of the solar photospheric abundance (the errors are at 90\% confidence level) versus the mean MOS count rate (in bin of $10''$) for the 16 subregions of Fig. \ref{fig:EPIC03-2_avgE}. The $Fe$ abundances can be deduced from the condition $(Ne/Ne_\odot)/(Fe/Fe_\odot)=4.4$. Cross$\equiv$FilD subregions; diamonds$\equiv$southwestern subregions; squares$\equiv$RegNE and North subregions.}
\label{fig:Ne}
\end{figure}

\begin{figure}[htb!]
\caption{Best fit values of $T_{I}$ (cooler component) e $T_{II}$ (hotter component) versus the mean MOS count rate (in bin of $10''$) for the 16 subregions of Fig. \ref{fig:EPIC03-2_avgE}. In subregions 14 and 15 (whose spectra are well described by one component) only $T_{II}$ is indicated. $T$ errors are at 90\% confidence level, while the maximum errors in the count rate are of $3\times 10^{-6}$ s$^{-1}$. Symbols as in Fig. \ref{fig:Ne}.}
\label{fig:T1T2}
\end{figure}

\begin{figure}[htb!]
 \centerline{\hbox{     
  }}
\caption{Best fit values of the emission measure per unit area (that is, the product of the square of the density $n$ of the plasma times its geometric extension along the line of sight $L$) of the cooler component, \emph{left panel}, and of the hotter component, \emph{right panel}, vs. the mean MOS count rate (in bin of $10''$) for the 16 subregions of Fig. \ref{fig:EPIC03-2_avgE}. For subregions 14 and 15 (whose spectra are well described only by the hotter component), an upper limit to the cooler component is reported. Errors as in \ref{fig:T1T2}, symbols as in Fig. \ref{fig:Ne}. FilD $\equiv$ subregions $1-6$; S-W $\equiv$ subregions $7-12$;  RegNE and North region $\equiv$ subregions $14-16$; the subregion 13 is between FilD and RegNE.}.   
\label{fig:N1-N2} 
\end{figure}

\begin{figure}[ht!]
\caption{3D map of the observed ISM clouds. The height of the features is proportional to the extension along the line of sight of the cores ($L_{I}$, in white) and of the coronae ($L_{II}$, semi-transparent blue). North and East are also indicated. In this image the structure of FilD is visible; the optical filament of Fig. \ref{fig:ottico_0305} lies between the two peaks, corresponding to subregions 2 and 4 of Fig. \ref{fig:EPIC03-2_avgE}.  Assuming $n_{I}=1.5$ cm$^{-3}$ and $n_{II}=0.55$ cm$^{-3}$, the maximum value for $L_{I}$ is $\sim 4.4\times 10^{18}$ cm and for $L_{II}$ is $\sim 4.7\times 10^{18}$ cm (for other values of density $L$ scale as $1/n^{2}$).}
\label{fig:3D1}
\end{figure}

\begin{figure}[ht!]
\caption{Same as Fig. \ref{fig:3D1} from a different angle-shot. The large corona of RegNE is visible in close-up.}
\label{fig:3D2}
\end{figure}

  \begin{figure}[tbh!]
   \centering
      \caption{Schematic representation of the structure of the FilD cloud. We report the values of the particle density $n$ and of the temperature $T$ of the plasma; the values of the optical filament temperature and density are taken from \citet{bms00}. The fat arrows indicate the evaporation of the plasma.}
         \label{fig:cloud}
   \end{figure}

\newpage

\appendix 

\section{Influence of the background on the spectral best-fit parameters}
 
As explained in Sect. \ref{Background Subtraction}, we used the EPIC public observation $0106260201$ to extract a pn and a MOS background spectrum for the spectral analysis. 

In Table \ref{tab:rxjgps} we report the best fit parameters obtained from the spectral analysis of subregions 4, 10, and 16, with the subtraction from source spectra of the background of the EPIC observation $0106260201$ and of the EPIC observations of the Galactic Plane Survey\footnote{10 EPIC observations of the Galactic Plane Survey, pointing galactic longitude $319^\circ$, and galactic latitude between $0^\circ$ and $3^\circ$.}. The table shows that the choice of the photon background does not significatively influence the values of the best-fit parameters. 

\begin{center}
\begin{table}[h!]
\begin{center}
\caption{Best fit parameters, obtained from the spectral analysis of subregions 4, 10, and 16 of Fig. \ref{fig:EPIC03-2_avgE} with subtraction from source spectra of the background of the EPIC observation $0106260201$ (left column) and of the background of the 10 EPIC observations of the Galactic Plane Survey (GPS, right column).}
\begin{tabular}{|l|c|c|} \hline\hline 
                                         &                              &                          \\ 
 \textbf{subregion 4}                    &         $0106260201$         &         GPS              \\ 
                                         &                              &                          \\ \hline
$T_{I}$ ($10^{6}$ K)                     &  $1.14^{+0.05}_{-0.04}$      & $1.15^{+0.04}_{-0.05}$   \\ \hline
$T_{II}$ ($10^{6}$ K)                    &       $3.4\pm 0.5$           &   $3.5^{+0.8}_{-0.6}$    \\ \hline
$n^{2}_{I}L_{I}$ ($10^{18}$ cm$^{-5}$)   &      $7^{+1.2}_{-1.4}$       &   $7^{+1.2}_{-1.1}$      \\ \hline
$n^{2}_{II}L_{II}$ ($10^{18}$ cm$^{-5}$) &    $0.28^{+0.09}_{-0.07}$    &  $0.28^{+0.09}_{-0.1}$   \\ \hline
$Ne/Ne_\odot$                            &     $1.3^{+0.7}_{-0.5}$      &    $1.3^{+1}_{-0.6}$     \\ \hline
$\chi^{2}$ (d. o. f.)                    &       102.1 (85)             &       93.0 (85)          \\ \hline
                                         &                              &                          \\ 
 \textbf{subregion 10}                   &      $0106260201$            &          GPS             \\ 
                                         &                              &                          \\ \hline
$T_{I}$ ($10^{6}$ K)                     &  $1.05^{+0.04}_{-0.06}$      & $1.01^{+0.07}_{-0.08}$   \\ \hline
$T_{II}$ ($10^{6}$ K)                    &    $3.3^{+0.2}_{-0.3}$       &   $3.1^{+0.3}_{-0.1}$    \\ \hline
$n^{2}_{I}L_{I}$ ($10^{18}$ cm$^{-5}$)   &      $4.5^{+0.8}_{-0.7}$     &   $5.1^{+1.5}_{-0.9}$    \\ \hline
$n^{2}_{II}L_{II}$ ($10^{18}$ cm$^{-5}$) &    $0.52^{+0.05}_{-0.04}$    &  $0.54^{+0.04}_{-0.06}$  \\ \hline
$Ne/Ne_\odot$                            &     $1.2^{+0.4}_{-0.3}$      &    $1.4^{+0.4}_{-0.2}$   \\ \hline
$\chi^{2}$ (d. o. f.)                    &       181.0 (157)            &       176.9 (157)         \\ \hline
                                         &                              &                          \\ 
 \textbf{subregion 16}                   &      $0106260201$            &          GPS             \\ 
                                         &                              &                          \\ \hline
$T_{I}$ ($10^{6}$ K)                     &       $1.3\pm 0.2$           &  $1.2^{+0.1}_{-0.6}$     \\ \hline
$T_{II}$ ($10^{6}$ K)                    &       $2.9\pm 0.2$           &   $2.8^{+0.2}_{-0.4}$    \\ \hline
$n^{2}_{I}L_{I}$ ($10^{18}$ cm$^{-5}$)   &      $0.16^{+0.6}_{-0.03}$   &  $0.19^{+0.15}_{-0.05}$  \\ \hline
$n^{2}_{II}L_{II}$ ($10^{18}$ cm$^{-5}$) &       $0.14\pm 0.01$         &  $0.14^{+0.03}_{-0.01}$  \\ \hline
$Ne/Ne_\odot$                            &     $2.0^{+0.5}_{-0.2}$      &    $2.2^{+0.5}_{-0.4}$   \\ \hline
$\chi^{2}$ (d. o. f.)                    &       189.1 (169)            &       185.5 (169)        \\ \hline
\end{tabular}
\end{center}
\label{tab:rxjgps}
\end{table}
\end{center}

\end{document}